    \documentstyle[nato,psfig]{crckapb}

\begin{opening}
\title{MIXING ALONG THE RED GIANT BRANCH IN \protect\\METAL-POOR FIELD STARS}

\author{E. CARRETTA$^1$}
\author{R.G. GRATTON$^1$}
\institute{Osservatorio Astronomico di Padova\\
           Vicolo dell'Osservatorio 5, I-35122 Padova, ITALY}
\author{C. SNEDEN$^2$}
\institute{Department of Astronomy and McDonald Observatory\\
           The University of Texas at Austin, USA}
\author{A. BRAGAGLIA$^3$}
\institute{Osservatorio Astronomico di Bologna\\
           via Ranzani 1, I-40127  Bologna, ITALY}

\end{opening}

\runningtitle{MIXING ALONG THE RGB}

\begin{document}


\begin{figure}
\vspace{7cm}
\includegraphics{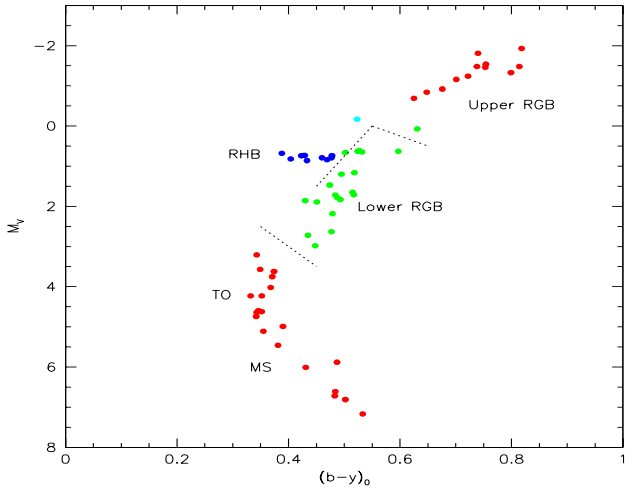}
\caption{Selected sample: we define as lower red giant branch (lower-RGB) the
evolutionary phase between first dredge-up completion and the RGB bump:
small mass stars on the lower-RGB
should have a well defined set of CNO and Li abundances, distinct from that
observed in main sequence stars (MS stars, which have
not yet experienced the first dredge up) and upper-RGB stars
(hereinafter, upper-RGB stars are those stars first ascending the
red giant branch, which are brighter than the RGB bump).}
\label{f:fig1}
\end{figure}

\section{Introduction}

Stellar models predict that as a small mass star evolves up the RGB, the
outer convective envelope expands inward and penetrates into the CN-cycle
processed interior regions ({\it first dredge-up}).
Approximately, the outer 50\% of the star by mass is involved in this 
mixing that brings to the surface mainly $^{13}$C and $^{14}$N, while
the primordial $^{12}$C and fragile, light elements like Li, Be, and B are 
transported from the surface to the interior.
Results for old disk field giants (Shetrone et al. 1993) and metal-poor stars
(Sneden et al. 1986) show that the first dredge-up occurs at the predicted
luminosities; however, mixing in bright giants is much more extreme than
predicted by evolutionary models.

On the other side, some further mixing is possible in the latest phases of the
RGB (see e.g., Charbonnel 1994, 1995): in fact,
after the end of the dredge-up phase is reached, the convective envelope
begins to recede, leaving behind a chemical discontinuity. This is
subsequently contacted by the H-burning shell, giving rise to the so-called
RGB bump.
Before this contact is made, the H-burning shell is advancing through a region
where there are appreciable composition gradients, which should inhibit any
(rotational induced) mixing.
Thereafter, since there is no mean molecular weight gradient between the
convective envelope and the near vicinity of the shell, it is possible that
circulation currents, perhaps driven by meridional circulations activated by
core rotation (Sweigart \& Mengel 1979) give rise to further mixing.

In principle, globular clusters offer a unique
opportunity to verify this scheme, since they provide a large number of
stars located at the same distance (thus allowing an accurate definition of the
evolutionary status of individual stars), having the same age (thus similar
masses evolving along the RGB), and, hopefully, the same initial chemical
composition. 
However, surface abundances of globular cluster RGB
stars have revealed a complex phenomenology (for a summary, see Kraft
1994), that has
defied insofar any attempt of a detailed explanation. This is likely due to the
fact that the surface abundances of these stars are significantly affected by
several major factors: deep mixing within individual stars,
primordial inhomogeneities within a cluster, and perhaps accretion of
nuclearly processed material during the early phases of the cluster evolution.
Furthermore, it is becoming increasingly clear that the dense environment plays
an important role on determining other basic cluster features
(like the colour of the horizontal branch), either by causing systematic
variations in the basic stellar properties (like e.g., the initial angular
momentum), or in favouring pollution of the surface layers of stars by ejecta
from other stars, or both.

To solve these issues, it is necessary to first understand the
evolution of single undisturbed small mass stars in the field, in a restricted
range of mass and metal abundance (both of them affecting the luminosity of
the RGB bump).

\section{Sample selection and observations}

To this purpose, we selected a sample from Anthony-Twarog \& Twarog (1994, ATT) for
the evolved stars) and Schuster \& Nissen (1989) for the main sequence and
turn-off stars, plus a few local subdwarfs from Clementini et al. (1998). 
We restricted to metallicities in the range 
$-2<$[Fe/H]$<-1$, in order to: (i) avoid possible massive interlopers (thin
disk stars), (ii) have a large number of moderately bright stars with accurate
absolute magnitudes, hence clear evolutionary phases, and (iii)  avoid
complications from large star-to-star abundance variations present among the
most metal-poor stars.

The resulting sample (62 stars) is shown in Figure~\ref{f:fig1};
evolutionary phases are derived from the position on the $(b-y)_0-c_{10}$ and 
the calassical colour magnitude diagram; M$_V$ values are from ATT (giants and 
subgiants) or from Hipparcos (dwarfs).

High S/N ($>100$), high resolution ($R>50,000$) spectra were acquired at the
McDonald Observatory (2.7m telescope $+$ ``2d-coud\'e'' echelle
spectrometer, spectral range 3800$\leq \lambda \leq$9000 ~\AA) and
at ESO (CAT$+$CES spectrograph, spectral regions centered at
4230, 4380, 5680, 6300, 6700, and 7780 ~\AA), to measure abundance indicators
for Li, C, N, O, Na and Fe.

\begin{figure}
\vspace{10cm}
\includegraphics{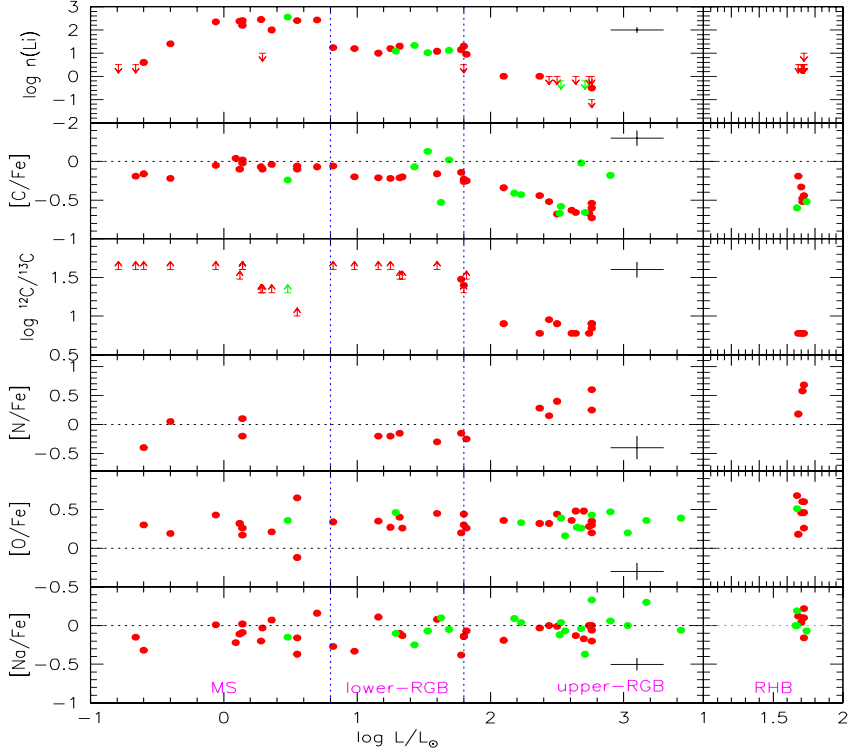}
\caption{Run of the abundance of Li, of the abundance ratios [C/Fe], [N/Fe],
[O/Fe] and [Na/Fe], and of the isotopic ratio $^{12}$C/$^{13}$C with luminosity
for stars with $-2<$[Fe/H]$<-1$. Filled symbols are actual measures; arrows
are upper (for Li) or lower (for $^{12}$C/$^{13}$C) limits. Typical
error bars for the various quantities are shown. Dashed
lines separate various evolutionary phases.
Results for red HB (RHB) stars are plotted separately for clarity.}
\label{f:fig2}
\end{figure}

\section{Atmospheric parameters and abundance analysis}

Effective temperatures (T$_{\rm eff}$'s) were derived from the
dereddened $(B-V)_0$
and $(b-y)_0$ colours using colour-T$_{\rm eff}$ transformations from Kurucz
(1995) models with no overshooting, and reddenings from ATT or Carretta,
Gratton \& Sneden (1999). Note that no empirical correction was required for
these calibrations (see e.g., Castelli, Gratton \& Kurucz 1997).
Gravities were derived from absolute magnitudes and T$_{\rm eff}$'s, 
assuming a mass M = 0.85 M$_\odot$ for all stars and B.C.'s from Kurucz (1995).
Eliminating trends of abundances derived from Fe I lines versus EWs for stars 
with McDonald
spectra, we obtained microturbulent velocities $v_t$'s for this subsample and a
tight relation $v_t=f$(T$_{\rm eff}$,log g). This was used to derive accurate
$v_t$ values for stars with CAT spectra (having very few Fe lines measured).

The abundance analysis was performed using Kurucz (1995) model atmospheres
with no overshooting. Further details are in Gratton et al. (2000).

Abundances for Fe I, Fe II, O I, Na I, $\alpha-$elements and heavy
elements were derived from measured equivalent widths (EWs).
Whenever possible, we complemented the rather few atomic lines measured on CAT
spectra with lists of accurate (errors $\leq 3$ m\AA) and homogeneous EWs from
literature (see Gratton, Carretta \& Castelli 1997 for references).

O abundances were obtained from both forbidden and permitted lines. Abundances
from these last (as well as the Na abundances from the 5682-88~\AA\ and
6154-60~\AA\ doublets) include corrections for departures from LTE 
(Gratton et al. 1999).
Average [O/Fe] ratios were computed after a small offset of 0.08 dex between
the abundances from [O I] and O I lines was accounted for.

C abundances as well $^{12}$C/$^{13}$C isotopic ratios were derived from
spectral synthesis of the G-band around 4300 \AA. Synthetic spectra
computations of the (0.0) and (1,1) bandheads of the violet band of CN were
used to derive N abundances for the 24 program stars with McDonald spectra.

Li abundances have been derived from comparison to
synthetic spectra, and complemented
with data from Pilachowski, Sneden \& Booth (1993, PSB), after Fe
and Li abundances from their analysis were increased to compensate the offsets
(0.09 and 0.21 dex, respectively) with respect to our results, due to our higher
T$_{\rm eff}$'s ($112 \pm 32$ K, from 14 stars in common with PSB).

\begin{figure}
\vspace{6cm}
\includegraphics{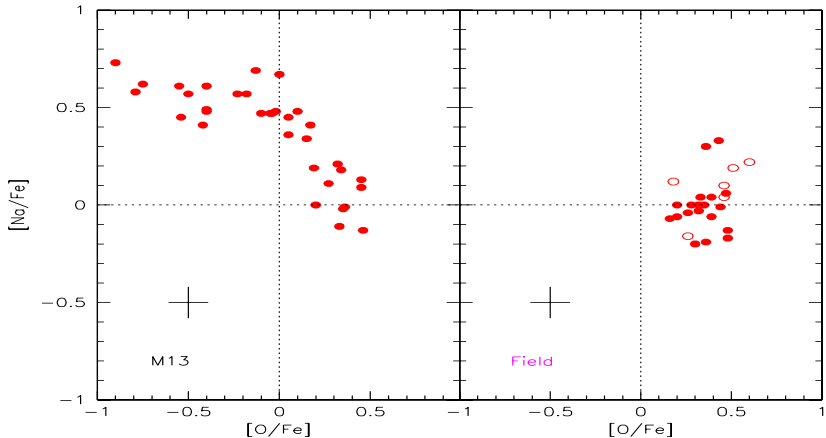}
\caption{The Na-O anticorrelation for field stars and for stars in M13.
Filled symbols are stars first ascending the RGB (only upper-RGB stars are
shown); open symbols are RHB stars. Error bars are shown at the bottom left.}
\label{f:fig3}
\end{figure}

\section{Results}

Results for Li, C, N, O, Na and $^{12}$C/$^{13}$C isotopic ratios are
summarized in Figure~\ref{f:fig2} for the 62 program stars with accurate
evolutionary phase, plus 43 stars more having accurate abundances for some of
these elements and similarly well defined luminosities from the literature.
>From Figure 2 we can see that small mass lower-RGB stars (i.e., stars
brighter than the first dredge-up luminosity and fainter than the RGB
bump) have abundances of light elements in agreement with predictions from
classical evolutionary models: only marginal changes occur for CNO elements,
while dilution within the convective envelope causes the surface Li abundance
to decrease by a factor of $\sim 20$.

A second, distinct mixing episode occurs in most (perhaps all) small mass
metal-poor stars just after the RGB bump, when the molecular weight barrier
left by the maximum inward penetration of the convective shell is canceled by
the outward expansion of the H-burning shell, in agreement with recent
theoretical predictions.

In field stars, this second mixing episode only reaches regions of
incomplete CNO burning: it causes a depletion of the surface $^{12}$C
abundance by about a factor of 2.5, and a corresponding increase in the N
abundance by about a factor of 4. The $^{12}$C/$^{13}$C is lowered to about 6
to 10 (close to, but distinctly larger than the equilibrium value of 3.5),
while practically all remaining Li is burnt.

However, an O-Na anti-correlation such as
typically observed among globular cluster stars (see Figure~\ref{f:fig3})
is not present in field stars. None of the 29 field stars more evolved than the
RGB bump (including 8 RHB stars) shows any sign of O depletion or Na
enhancement. This means that in field stars the second mixing episode is not
deep enough to reach regions were ON-burning occurs.

\end{document}